# Insights into Formation of Bicontinuous Emulsion Gels via *in-situ* (Ultra-)Small Angle X-ray Scattering

*Meyer T. Alting[1], Dominique M.E. Thies-Weesie[1], Alexander M. van Silfhout[1,2], Mariska de Ruiter[1], Theyencheri Narayanan[3], Martin F. Haase[1*], Andrei V. Petukhov[1*]*

[1] Van 't Hoff Laboratory for Physical and Colloid Chemistry, Department of Chemistry, Debye Institute for Nanomaterials Science, Utrecht University, Utrecht, The Netherlands

[2] *Current affiliation*: TNO Environmental Modelling, Sensing and Analysis, Princetonlaan 6, 3584CB Utrecht, The Netherlands

[3] ESRF – The European Synchrotron, 71 Avenue des Martyrs, 38043 Grenoble, France

* Correspondence: a.v.petukhov@uu.nl and m.f.haase@uu.nl

**Abstract**

Nanostructured materials formed via kinetically controlled self-assembly processes gather more interest nowadays. Bicontinuous emulsion gels stabilized by colloidal particles, called bijels, are attractive materials in soft-matter as they combine bulk properties of two immiscible liquids into an interwoven network structure. The limited understanding of the complex formation phenomena of bijels restricts the control over the synthesis, and so its applicability. In this work, *in-situ* (ultra-) small-angle X-ray scattering is applied to gain insight into the phase separation and self-assembly kinetics of bijels formed via solvent transfer induced phase separation. An X-ray compatible microfluidic setup allows accessing the process kinetics with a millisecond resolution. The formation of such bijels is shown to occur via three consecutive steps related to fluid mechanics, nanoparticle self-assembly and liquid-liquid phase separation. This time-resolved monitoring technique offers valuable insights into the structural evolution of kinetically controlled materials and enhances our understanding of the formation of bicontinuous emulsion gels.

Kinetically controlled self-assembly of nanostructured materials is an active research field due to its diverse applicability and complex formation mechanisms.[1–4] Since two decades, fluid bicontinuous emulsions gels (bijels) make their advance in this field of self-assembly.[5–7] Bijels consist of two immiscible liquids that are mechanically arrested in a non-equilibrium, bicontinuous arrangement by a percolating layer of colloids at the liquid-liquid interface. These unique structures have high potential in areas like catalysis[8,9], separation membranes[10], biomaterials[11,12] and energy storage[13,14]. Limited control over the formation mechanisms of bijels, however, restricts their applicability.

The formation of bijels is a complex, non-equilibrium self-assembly process.[7,15] Generally, bijels are formed by inducing phase separation of a liquid mixture containing surface-active colloidal particles into two immiscible liquids via spinodal decomposition.[6,16–20] The particles self-assemble on the interface between both interwoven liquids to minimize the interfacial area. Eventually, the particles form a dense layer around the liquid domains that kinetically arrests the liquids from further thermodynamically-driven de-mixing. The obtained structure forms a viscoelastic bicontinuous emulsion with fully interconnected liquid domains.[21]

The obtained structure of the bijel sensitively depends on many factors, such as the exact formulation of the precursor mixture, type of colloidal particles and their surface chemistry, temperature and fabrication method.[14,19,22–24] To the best of our knowledge, studies on the influence of these factors are based on *ex-situ* characterization of bijel structures *after* completing phase separation. The phenomena occurring *during* spinodal decomposition such as the self-assembly of particles have not yet been extensively investigated *in-situ* due to common limitations in microscopy like optical transparency or lack of contrast.[6] Prior research on bijel-templated membranes proposed a method to study bijel formation by polymerizing intermediate phase separated stages.[10,25] Since this method requires polymerization of one liquid phase, restricting its applicability to such polymerizable bicontinuous emulsions. The complex control over the formation of bijels consisting of two liquid phases calls for new *in-situ* methods to study the phase separation and consecutive self-assembly of particles *during* phase separation.

Here, we present a microfluidic setup suitable for synchrotron small angle X-ray scattering (SAXS) to investigate the dynamic self-assembly of colloids during bijel formation in real time. The advantage of SAXS over microscopy techniques is that it neither requires optical transparency nor changes to experimental conditions: nearly all nanoparticle and colloidal systems in both soft- and hard matter can be probed.[26–31] *In-situ* SAXS has already been successfully employed in monitoring self-assembly behavior of nanoparticles in the formation of e.g. emulsions, mesoporous materials and nanostructures.[32–35] Traditional SAXS gives access to structural features on the nanometer scale and is suitable to observe primary particles and their positional correlations with neighboring particles during phase separation (~$10^0 - 10^2$ nm). Extending the technique to ultra-small angles (USAXS) enables analysis on larger scales (~$10^2$ – $10^3$ nm) to shed light onto larger features formed during phase separation like aggregates and pores.[36–39]

In this work, we investigate the kinetics of phase separation and self-assembly of nanoparticles during bijel formation. We perform *in-situ* SAXS measurements during the synthesis of pinched-off bijel fragments fabricated out of a critical mixture undergoing spinodal decomposition via solvent transfer induced phase separation (STrIPS) in a microfluidic device. SAXS analysis reveals that individual nanoparticles become more attractive and self-assemble at the liquid-liquid interface during phase separation. We propose a two-phase theoretical model to simulate the attachment of the particles onto the interface. Furthermore, we show that USAXS probes the phase separation itself and the formation of the tortuous structure. In addition, the combination of SAXS and USAXS reveals that the particle self-assembly occurs significantly earlier than the spinodal decomposition itself. These results show that the self-assembly of the particles and the formation of the tortuous bijel network can be followed in real time. This time-resolved methodology can be extended to other fabrication methods and critical mixtures to obtain a deeper understanding of the kinetic pathways involved during bijel synthesis.

In this paper, we focus on the formation of bijels with a interwoven submicron network interior. **Figure 1A** shows a cross-section of a bijel fiber imaged by confocal laser scanning microscopy (CLSM) and scanning electron microscopy (SEM). CLSM identifies the interwoven oil- (labeled magenta) and aqueous networks (labeled black) stabilized by a rigid particle layer (labeled green). SEM shows the range of pore sizes between 200 nm near the outer surface to several microns near the center. This pore size gradient evolves due to the formation mechanism of bijels as reported previously.[40,41] Recent work showed that these bijels can be fabricated as fibers from a critical precursor mixture via solvent transfer induced phase separation (STrIPS).[40–42] This precursor mixture contains oil (diethyl phthalate), aqueous (water and glycerol), solvent (1-propanol), nanoparticles (silica Ludox TMA, 29 nm) and surfactant (cetyltrimethylammonium bromide, CTAB) (see **Figure 1B)**. During STrIPS, the precursor mixture is injected into toluene. 1-propanol diffuses from the precursor into toluene and triggers phase separation. Consequently, CTAB-modified silica nanoparticles self-assemble on the liquid-liquid interface and arrest the bicontinuous structure as schematically depicted in **Figure 1C**.

To study nanoparticle self-assembly during phase separation, *in-situ* SAXS and USAXS experiments are performed at the ID02 beamline at the European Synchrotron Radiation Facility (ESRF), Grenoble, France (see SI S1.7).[36] A microfluidic device consisting of two coaxially aligned glass capillaries is horizontally positioned at the beamline, as illustrated in **Figure 1D** (see SI S1.3 and S1.4). Bijels are continuously extruded as pinched-off fragments by injecting a precursor mixture via an inner capillary into a flow of toluene in a outer capillary (see SI S1.5 and S2).[43] While traversing through toluene, the fragments undergo STrIPS and form phase separated bijel structures. Measuring the scattering of fragments at various positions along the capillary enables *in-situ* tracking of the transition from a homogeneous precursor mixture into phase separated structures. The positions are converted into time scales using the estimated velocity of the fragments through the capillary based on the flow rates of the precursor and toluene (see SI S2 and S3). These measurements provide time-resolved insights into phase separation and self-assembly.

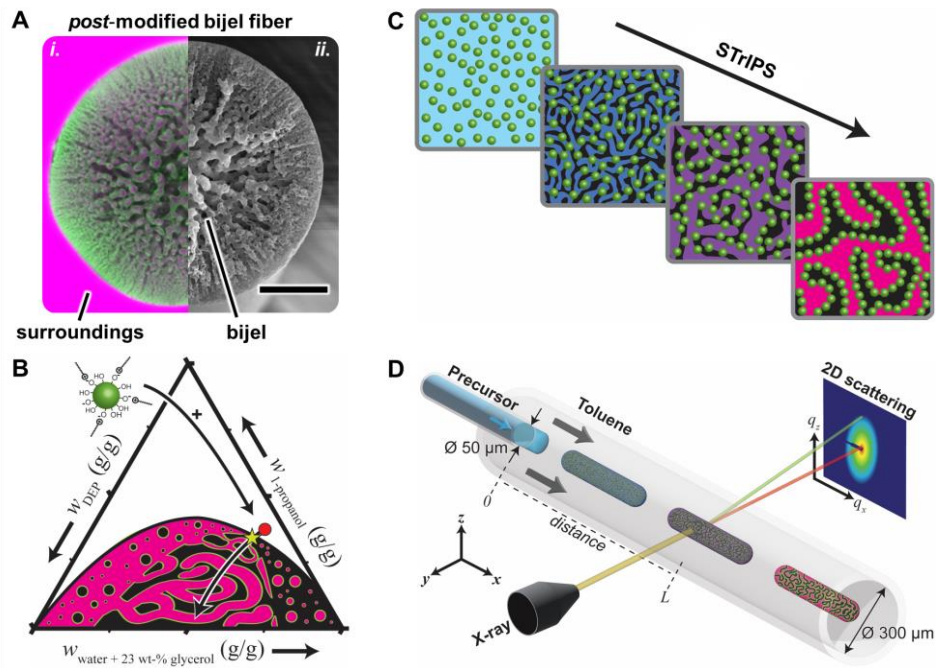

**Figure 1**: <u>Bijel formation.</u> **A)** Cross-section of bijel fiber imaged by **i)** confocal- and **ii)** scanning electron microscopy. Magenta is oil, green is particles and black is aqueous phase. Scale bar = 20 µm. **B)** Phase diagram of diethyl phthalate (DEP), water, glycerol and 1-propanol as weight fraction $w_i$ with binodal curve, critical point (yellow star) and precursor liquid composition with CTAB-modified Ludox TMA particles (red dot). Illustration below the binodal represents the regions of spinodal decomposition and nucleation. **C)** Schematic bijel formation via spinodal decomposition and kinetic arrest of liquid domains. **D)** *In-situ* microfluidic extrusion device for *in-situ* measurements at a beamline.

In the following, we first explore the self-assembly kinetics on the nanoparticle level during phase separation as acquired by SAXS. Thereafter, we extend the discussion on the structural evolution to larger scales like bicontinuous network formation as probed by USAXS. Unless stated otherwise, the scattering intensity profiles presented in this paper have been normalized and background-corrected according to standard procedures (see SI S4 and S5).[36]

**Nanoparticle assembly during phase separation**. We first study the X-ray scattering on the nanoparticle level when fragments undergo phase separation. **Figure 2A** shows the scattering intensity profiles $I(q, t_i)$ against the scattering vector $q$, recorded *in-situ* during the formation of a bijel fragment for time $t_i$ from 0 to 611 ms (see SI S6). Here, the $q$-value ranges from the greatest of 1.0 nm$^{-1}$ and 0.025 nm$^{-1}$ as smallest, corresponding to spatial scales around, respectively, 6 and 250 nm. In this $q$-range, the patterns are dominated by the scattering of silica nanoparticles (see SI S11).[44] For $q > 0.30$ nm$^{-1}$, $I(q, t_i)$ originates from the scattering of individual silica particles with fringes characteristic for spherical particles.[45] Thus, to observe the particle organization and to reveal their spatial correlations during phase separation, one has to focus on the region below $q = 0.30$ nm$^{-1}$.

In the region for $q < 0.30$ nm$^{-1}$, two iso-scattering points in $I(q, t_i)$ can be observed at $q_1 = 0.063$ and $q_2 = 0.226$ nm$^{-1}$. In between these iso-scattering points, $I(q, t_i)$ decreases over time. Simultaneously, the region below $q_1 = 0.063$ nm$^{-1}$ shows a strong upturn in intensity. These variations in $I(q, t_i)$ suggest that the positional correlation between nanoparticles changes over time, as discussed next.

We point out that $I(q, t_i)$ depends on structural interactions between the particles. The scattering intensity $I(q)$ can be factorized into the product of a form factor $P(q)$ and structure factor $S(q)$ via $I(q) \propto P(q)S(q)$. Here, $P(q)$ describes the scattering of individual silica nanoparticles (SI S7). $S(q)$ depends on the positional correlations between particles and can be extracted by dividing $I(q)$ by $P(q)$, followed by normalizing the high $q$ limit to 1 (see SI S8).

**Figure 2B** shows the structure factor $S(q, t_i)$ calculated from the scattering intensity profiles $I(q, t_i)$ for $t_i$ between 0 and 611 ms (see SI S8). Initially, $S(q, t_i)$ exhibits only slight variations as a function of $q$ as expected for a dilute suspension. Over time, three main features in $S(q, t_i)$ can be observed: I) a maximum develops at $q_{max} = 0.26$ nm$^{-1}$, II) $S(q, t_i)$ reduces between the iso-scattering points, and III) $S(q, t_i)$ sharply increases for the region below iso-scattering point $q_1$.

The first two changes mentioned provide structural insights into the self-assembly of the nanoparticles over time. The correlation peak $S(q_{max} = 0.26 \text{ nm}^{-1}, t_i)$ indicates the mean core-to-core distance between particles (see SI S8).[46] In real space, this peak corresponds to 24 nm, similar to the diameter of the nanoparticles (29 nm). Besides, the reduction in $S(q, t_i)$ between the iso-scattering points indicates less compressibility in the continuum limit, which is characteristic for the formation of densely packed structures. Together, these two effects suggest that nanoparticles self-assemble from a homogeneous dispersion into closely packed structures during phase separation. The third change in $S(q, t_i)$ i.e. the sharp increase for $q < 0.063$ nm$^{-1}$, indicates that the closely packed particles are organized into larger structures during the self-assembly. This intensity upturn can be used to identify the onsets and kinetics of particle self-assembly and formation of larger structures.

To this end, $S(q, t_i)$ has been plotted against time at arbitrary $q$ values of 0.150 nm$^{-1}$ and 0.025 nm$^{-1}$, corresponding, respectively, to the self-assembly of individual nanoparticles and the formation of larger structures, see **Figure 2C**. This plot shows that both $S(q = 0.150 \text{ nm}^{-1}, t)$ and $S(q = 0.025 \text{ nm}^{-1}, t)$ profiles change monotonically from the beginning of the extrusion onwards. However, whereas $S(q = 0.150 \text{ nm}^{-1}, t)$ continuously decreases for the full 600 ms measured, $S(q = 0.025 \text{ nm}^{-1}, t)$ reaches a plateau from 350 ms and kept nearly constant at longer times. This difference between the time dependences of the structure factor is not clear yet and needs further investigation.

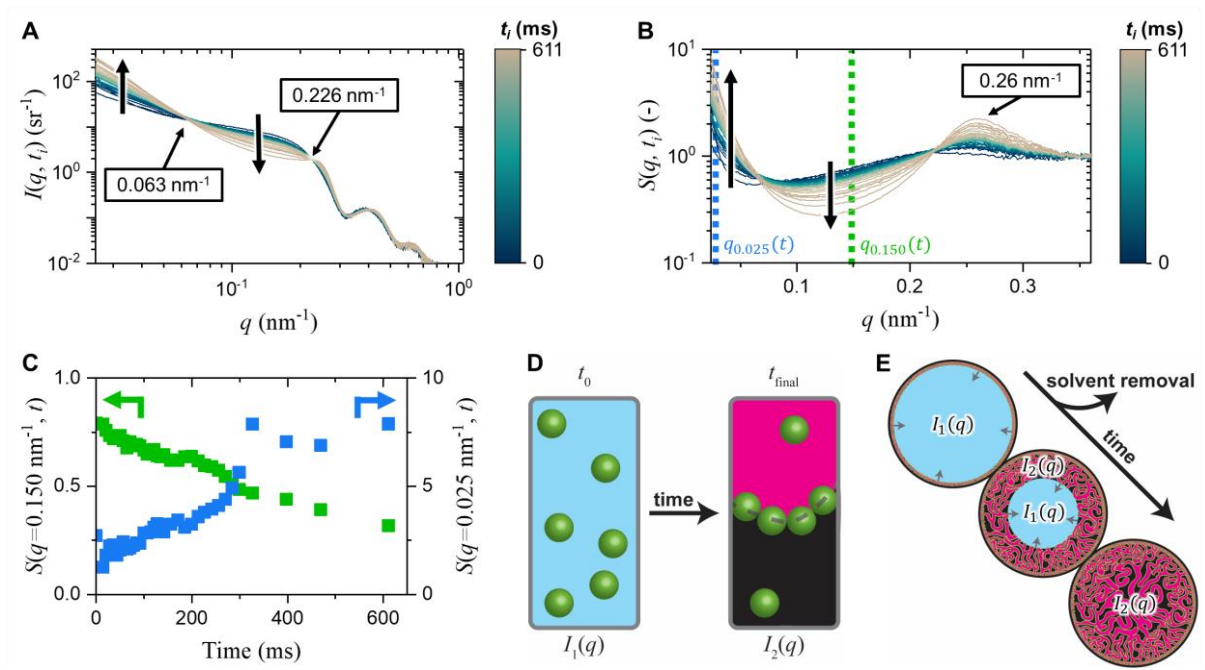

**Figure 2**: <u>Nanoparticle assembly during phase separation acquired by *in-situ* SAXS.</u> **A)** Normalized scattered intensity profiles on a double-log scale during bijel formation over time. **B)** Normalized time-resolved structure factors on a log-linear scale during bijel formation. **C)** Time-resolved structure factor at (green symbols, left axis) $q = 0.150$ nm$^{-1}$, nearest neighbor interactions, and (blue symbols, right axis) $q = 0.025$ nm$^{-1}$, larger structural contributions. **D)** Proposed two-state system on particle assembly from randomly dispersed particles with scattering intensity $I_1(q)$ with free nanoparticle volume fraction $\phi$ into an ordered layer of adsorbed particles at the oil-water interface with scattering intensity $I_2(q)$. **E)** Schematic representation of radially inward formation of bijel fragment during STrIPS and the contribution on the superposition of the scattering of $I_1(q)$ and $I_2(q)$.

The presence of the iso-scattering points in the scattering data suggests that particles are present in two competing states during phase separation. In this case, the scattering of one state, $I_1(q)$, would originate from single nanoparticles dispersed in a liquid medium with a flat structure factor. During phase separation, however, the nanoparticles start to adsorb at the interfaces with a condensed structure with a different $I_2(q)$ profile as illustrated in **Figure 2D**. Since the condensed phase grows at the cost of the dispersed phase, their contributions can be described by a single parameter $\phi$, the fraction of freely dispersed particles. As discussed in more detail in SI S9, the main features of the measured $I(q,t)$ dependence can be clarified within this simple model. The time dependence of the $\phi$ parameter can describe the radially inward propagation of the phase separation front from the outer surface of the fiber fragment towards its center as shown in **Figure 2E**.

In summary, the SAXS data indicate that the transition from a single phase with uniformly distributed silica nanoparticles, to their dense assemblies at the interfaces, takes at least 600 milliseconds. In addition, the strong intensity upturn at smaller $q$ indicates the formation of larger structures, which will be discussed in more detail below with the USAXS data.

**Bicontinuous network channel formation**. In the following, we will extend the discussion to ultra-small X-ray scattering (USAXS) to analyze scattering patterns for $q$-ranges corresponding to the micron scale (see SI S5). **Figure 3A** displays the evolution of the 2D scattering patterns for early stages of the STrIPS process up to 67 ms. Initially, we see strong isotropic scattering in the first few milliseconds after the extrusion. Surprisingly, most of this signal disappears for about ~50 ms. After this time, strong scattering starts to develop at $q$-values corresponding to submicron scales. The intensity of this scattering pattern grows in time and shifts simultaneously towards lower $q$-values. Remarkably, this scattering appears strongly anisotropic and is mostly observed along the fiber direction.

To better understand the evolution of these 2D scattering patterns, the 1D intensity scattering profiles around the $q_x$ axis were analyzed. **Figure 3B** shows $I(q, t_i)$ profiles averaged over an azimuthal angular range of $\pm 20°$ as shown in the inset of **Figure 3B** (see SI S5). Similar to SAXS data, the amount of material within the irradiated area fluctuates because of extrusion instabilities, undulation of the fiber, and so on. To remove these intensity fluctuations, the $I(q, t_i)$ profiles were normalized to the same intensity at the iso-scattering point at $q_1$ = 0.063 nm$^{-1}$ as earlier found by SAXS. For $q$ < 0.063 nm$^{-1}$, a strong scattering intensity can be seen in the first few milliseconds and after longer times in agreement with the 2D scattering patterns. However, the strong scattering signals in the first few ms are different than after 50 ms. The former is isotropic and exhibits no maximum. The latter is anisotropic and shows a broad maximum as a function of $q$.

To quantify the evolution of the scattering profiles over time, **Figure 3C** plots $I(q, t_i)$ against time at an arbitrarily selected $q$ = 0.011 nm$^{-1}$ with respect to $q_1$. This plot shows that three phases in $I(q = 0.011 \text{ nm}^{-1}, t)$ can be observed: I) a rapid decrease within 7 ms, II) a constant intensity till 50 ms, and III) a strong increase from 50 ms onwards.

In addition, **Figure 3D** plots the position, $q_{\max}$, of the broad maximum peak in phase III, against time and the corresponding distance, $d = 2\pi/q_{\max}$, in direct space. This plot shows that $q_{\max}$ lowers from $10 \cdot 10^{-3}$ to less than $8 \cdot 10^{-3}$ nm$^{-1}$, corresponding to an increase from 0.6 μm to over 0.8 μm in real space within 67 ms. No longer times have been measured for this extrusion due to practical constraints.

What does the evolution of the scattering profiles reveal about the synthesis of bijel fragments formed via STrIPS? The three distinct phases in $I(q = 0.011 \text{ nm}^{-1}, t)$ suggest that the formation occurs via three consecutive steps. To further elaborate on this question, we point out that origin of the scattering intensity measured in USAXS originates from both nanoparticles *and* phase separated liquids (see SI S11).

We begin the discussion of the USAXS starting from phase I, which corresponds to the injection of the precursor mixture into the flow of toluene. The strong scattering observed here is not present in a similar USAXS measurement performed under static conditions (see SI S10). This additional scattering signal must therefore originate either from the hydrodynamic mixing in the precursor as well as by the release of the shear stress induced by the inner capillary walls after the orifice (see SI S3).[47] As a result, the laminar velocity profile in the precursor has to quickly change due to the removal of the friction caused by the capillary walls. This change in the profile may affect the structural organization in the precursor.[43] However, the exact mechanism behind this transition requires further investigation.

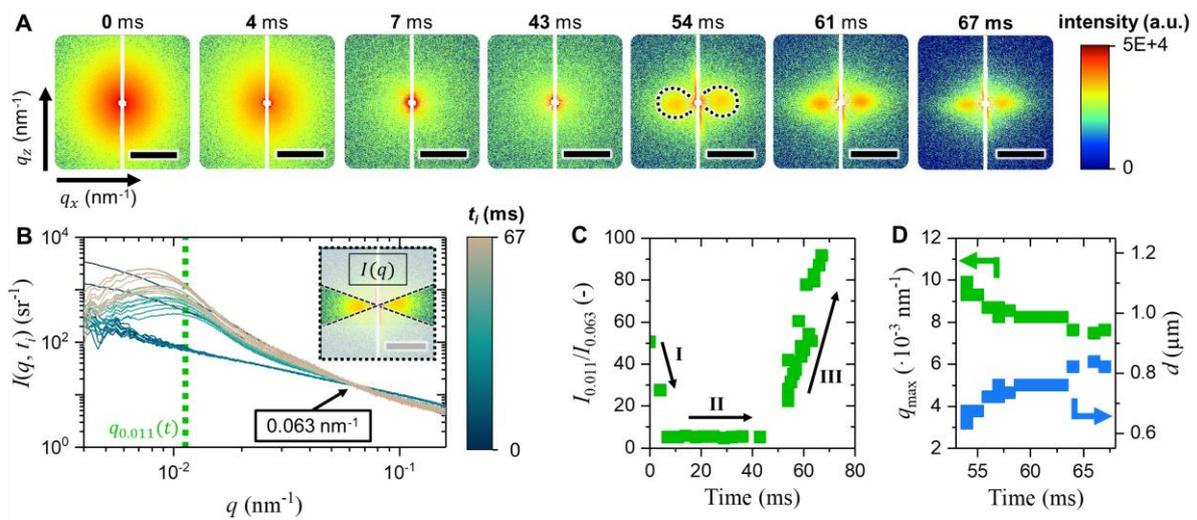

**Figure 3**: *In-situ* USAXS analysis on bijel formation. **A)** 2D scattering patterns of bijel fragments after various time steps. The dashed lines in the frame of 54 ms indicate the circumference of the lobs formed. **B)** Normalized scattering 1D intensity plots after averaging the bright region of the 2D scattering pattern shown in the inset. Normalization has been done at the iso-scattering point at $q_1 = 0.063$ nm$^{-1}$ as found by SAXS in Figure 2A. **C)** Time-dependence of the normalized intensity at $q = 0.011$ nm$^{-1}$ with phases I, II and III indicated. **D)** Time-dependence of position of the broad correlation maximum $q_{max}$ and corresponding characteristic interwall distance against STrIPS time determined from the structure factor (see SI S8).

Proceeding to phase II, the constant low values in $I(q,t)$ indicate that no larger structures are formed in the precursor. This suggests that the precursor did not undergo yet phase separation. We conclude that there is a particular waiting time before the concentration of 1-propanol sufficiently lowered to induce phase separation. In agreement with the phase diagram in **Figure 1B**, the precursor mixture contains a small excess of 1-propanol required to form a stable miscible mixture.

Finally, the strong increase in $I(q,t)$ in phase III indicates that larger structures are being formed. The correlation peak present in the scattering intensity indicates that a bicontinuous structure is formed via spinodal decomposition. The strongly anisotropic shape of the correlation peak suggests that the formed structures are influenced by the presence of a shear force[48–51], and the inward progression of the phase separation leading to the radial orientation of the pores in a bijel fiber as shown in **Figure 1A**.

Before concluding, we compare the kinetic phenomena involved in bijel formation via STrIPS of the precursor mixture used in this study. Previous work modeled the diffusion of 1-propanol out of the precursor mixture and found that the radially inward-directed phase separation begins from 20 ms after extrusion.[40] Here, USAXS indicated a waiting time of 50 ms before spinodal decomposition occurs, similar to the simulations. Meanwhile, SAXS revealed that some nanoparticles already self-assemble during the waiting period from 0 ms onwards. These results indicate that some particle re-arrangement and assembly already occurs from the beginning of the extrusion. However, the actual formation of a bijel requires a slight waiting time before forming a bicontinuous structure via spinodal decomposition.

To summarize, combined time-resolved SAXS and USAXS measurements allow real-time monitoring of the complex formation of bicontinuous emulsion gels, a process previously experimentally inaccessible. Synchrotron SAXS/USAXS enables precise characterization of liquid phase separation and colloidal self-assembly on a millisecond timescale. SAXS measurements capture the transition from individually dispersed particles to condensed self-assembled structures, while USAXS elucidates the formation of larger structures composed of both liquids and particles. For the first time, *in-situ* experimental evidence confirmed that bijels synthesized from a critical mixture undergoing solvent transfer induced phase separation (STrIPS) are formed via spinodal decomposition. In addition, we have identified that the formation of such bijels occurs via three consecutive steps: kinetic phenomena related to fluid mechanics, nanoparticle self-assembly and liquid phase separation, each occurring on different timescales. While the exact mechanisms of these steps remain unclear, it offers opportunities for further investigation. The findings already presented in this work provide new insights into the kinetically controlled formation of bicontinuous emulsion gels. It opens up new analysis methods to characterize the synthesis mechanisms of various complex nanostructures. Eventually, this research can improve the understanding of complex assembly phenomena and enhance the potential of future applications.

**Author contributions**

**Meyer T. Alting**: Conceptualization, methodology, investigation, visualization, data curation, writing original draft. **Dominique M.E. Thies-Weesie**: Investigation, data curation. **Alexander M. van Silfhout**: Data curation. **Mariska de Ruiter**: Conceptualization, data curation, investigation. **Theyencheri Narayanan**: Methodology, investigation, visualization, data curation. **Martin F. Haase**: Supervision, project administration, funding acquisition. **Andrei V. Petukhov**: Conceptualization, investigation, visualization, writing original draft.

**Notes**

The authors declare no competing financial interest.

**Acknowledgement**

The authors thank the European Synchrotron Radiation Facility (ESRF) for provision of synchrotron radiation facilities (proposals IH-SC-1791 and SC-5514). We would like to thank the technical team for ID02 for the wonderful assistance and support in using beamline ID02. This project has received funding from the European Research Council (ERC) under the European Union's Horizon 2020 research and innovation program (Grant agreement no. 802636).

**Keywords:** bijels, colloids, phase separation, self-assembly, (U)SAXS

**Graphical abstract**

Colloidal self-assembly into bicontinuous structures is gaining interest for applications like energy storage and catalysis. In this work, we demonstrate the potential of an *in-situ* SAXS/USAXS set-up to monitor the time-resolved formation of bicontinuous emulsions. The results shed light onto the self-assembly of particles and liquid-liquid phase separation.

**in-situ (U)SAXS on growth of bicontinuous emulsions**

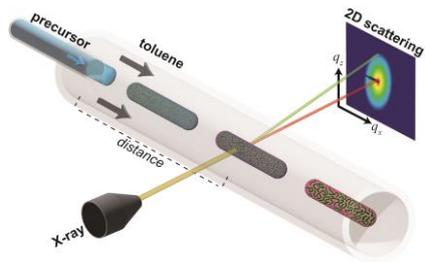

**Supporting Information**

**Insights into Formation of Bicontinuous Emulsion Gels via *in-situ* (Ultra-)Small Angle X-ray Scattering**

*Meyer T. Alting[1], Dominique M.E. Thies-Weesie[1], Alexander M. van Silfhout[1,2], Mariska de Ruiter[1], Theyencheri Narayanan[3], Martin F. Haase[1*], Andrei V. Petukhov[1*]*

[1] Van 't Hoff Laboratory for Physical and Colloid Chemistry, Department of Chemistry, Debye Institute for Nanomaterials Science, Utrecht University, Utrecht, The Netherlands

[2] *Current affiliation*: TNO Environmental Modelling, Sensing and Analysis, Princetonlaan 6, 3584CB Utrecht, The Netherlands

[3] ESRF – The European Synchrotron, 71 Avenue des Martyrs, 38043 Grenoble, France

\* Correspondence: a.petoukhov@uu.nl and m.f.haase@uu.nl

**Supporting Information 1          Experimental methods**

1.1 **Materials**

All chemicals were used as received. Diethyl phthalate (DEP, 99%), glycerol (99+%, synthetic), hydrochloric acid (37%) and toluene (99+%, extra pure) were received from Thermo Scientific. Hexadecyltrimethylammonium bromide (CTAB, ≥99%), light mineral oil (density 0.84 g mL$^{-1}$) and 1-propanol (≥99.5%) were purchased from Sigma-Aldrich. Silica nanoparticles (Ludox® TMA, batch number 1003481587, particle diameter 29 nm) were obtained from Grace GmbH. *n*-Hexane (99%, HPLC) was received from Biosolve BV. Octadecyl trichlorosilane (OTS, 94.3%) was purchased from Santa Cruz Biotechnology, Inc. Water used in all experiments is ultrapure MilliQ purified by a Rephile Genie U2 system with a resistivity of 18.2 MΩ·cm.

1.2 **Preparation of bijel precursor**

100 g of 34 wt-% Ludox TMA dispersion (batch number 1003481587) is concentrated from 34 wt-% to 52 wt-% in a rotary evaporator (Heidolph Instruments) at 60 °C and 140 mbar. The dispersion is centrifuged at 3750 g for 15 minutes (Allegra X-12R, Beckman Coulter) to remove any particle aggregates present. The weight percentage was determined by evaporation of 2 mL supernatant and consequent dry mass determination. Then, the dispersion was diluted to 50 wt-% by adding MilliQ. The particles were acidified to pH 1.8 using 1 M aqueous HCl shortly before adding to the precursor as described below.

The bijel precursor mixture consists of four liquids with weight fractions $w_{\text{DEP}} = 0.084$, $w_{\text{H}_2\text{O}} = 0.423$, $w_{\text{glycerol}} = 0.121$ and $w_{1-\text{propanol}} = 0.372$. In addition, the mixture contains Ludox TMA nanoparticles with a weight fraction of 0.301 and a CTAB concentration of 28.2 mM with respect to, respectively, the total mass and volume of the four liquids. 13 mL precursor is prepared by mixing the following liquids: 0.853 g DEP, 1.209 g of 200 mM CTAB in 1-propanol, 2.446 g of 50 wt-% glycerol in 1-propanol, 1.338 g 1-propanol and 8.650 g of the 50 wt-% Ludox TMA dispersion as discussed above. Additional details about the preparation of these bijel precursor mixtures can be found in references [1] and [2].

1.3 **Assembly of microfluidic device**

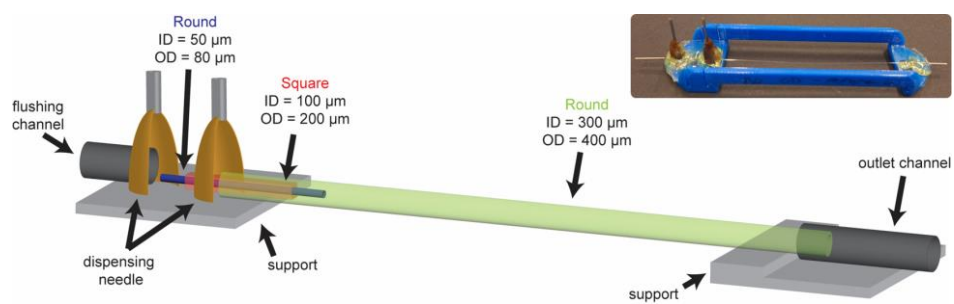

**Figure S1.3:** Schematic representation of microfluidic device for extruding bijel fragments. Photograph shows an actual device placed in a custom-made holder.

All borosilicate glass capillaries are purchased from CM Scientific. A round capillary (inner diameter (ID) 50 μm, outer diameter (OD) 80 μm, CV0508) is glued in a square capillary (ID 100 μm, OD 200 μm, 8510-050) with Norland adhesive 81. This assembly is inserted into a round capillary (ID 300 μm, OD 400 μm, CV3040) and centered. The ends of the square- and outer round capillaries are glued on separate microscopy slides using epoxy glue (Liqui Moly 6183). The assembly is glued on a custom made holder leaving the capillaries uncovered (**Figure S1.3**). Needles as flushing- and outlet channel are glued, respectively, near the 50 μm capillary and on the end of the outer capillary. Two dispensing needles are glued above the openings of both round capillaries as inlets. The needle-assembly is embedded in epoxy glue to prevent leakage.

### 1.4 Microfluidic device hydrophobization

A microfluidic device was initially filled with 9M aqueous KOH/NaOH solution for 6 hours and rinsed with water. Then, 3 vol-% OTS solution in mineral oil was filled in the capillaries, heated for 2 minutes using hot air and rinsed with *n*-hexane.

### 1.5 Enrichment of toluene by water

Toluene was enriched with water by vigorously shaking 250 mL toluene with 20 mL MilliQ for 5 min at room temperature. The mixture phase separated for 1 hour prior using the top-layer of the oil-phase.

### 1.6 Bijel fragment extrusion in microfluidic device in SAXS beamline

3 mL precursor mixture was loaded in a 5 mL syringe (BD Discardit™, Ø12.40 mm) and 20 mL water-enriched toluene in a 20 mL syringe (BD Discardit™, Ø20.00 mm). The syringes are connected to the inlets of the microfluidic device using PTFE tubing (Cole-Parmer Instrument Company). The outlet of the device is connected to a 100 mL collection vial using the same tubing. The syringes are placed in two syringe pumps (World Precision Instruments, model AL-1010). The microfluidic device is positioned horizontally on a movable stage in the beamline as depicted in **Figure 1D** in the main text. The precursor mixture flows with a flow rate of 0.40 mL h$^{-1}$ and, simultaneously, water-enriched toluene with a flow rate of 17.50 mL h$^{-1}$ through the microfluidic device. 50 μm diameter 5 mm long bijel fragments were extruded at ~30 s$^{-1}$ with a spacing of 3 – 5 mm between the fragments.

### 1.7 X-ray scattering experiments and data analysis

Static and time-resolved SAXS/USAXS measurements were performed at the ID02 time-resolved ultra small-angle X-ray scattering beamline at the European Synchrotron Research Facility (ESRF). The beam had an energy of 12.4 keV, corresponding to a wavelength of 0.1 nm and was 40 μm · 40 μm in size (FWHM). A Eiger2 4M detector was used for both SAXS and USAXS measurements by varying the sample-detector distance between 3 m (SAXS) and 31 m (USAXS). These distances cover a scattering vector $q$ between $1.73 \cdot 10^{-3}$ to 2.5 nm$^{-1}$ ($q = \frac{4\pi}{\lambda} \sin\left(\frac{\theta}{2}\right)$ with $\theta$ as scattering angle and $\lambda$ as wavelength).

Static measurements were conducted to characterize structure of silica Ludox TMA nanoparticles. Quartz capillaries (1.5 mm diameter, wall thickness 10.0 μm) were loaded by aqueous Ludox TMA dispersions at pH 1.8 with weight fractions ranging between $1 \cdot 10^{-3}$ and 0.50 g/g. The temperature in the experimental room was set to 23.5 °C. Ten frames with an exposure time of 0.10 s per frame were collected in both SAXS and USAXS.

Real-time scattering measurements on the extrusion of bijel fragment were performed by positioning the microfluidic device horizontally on a movable stage in the beamline. Different positions from the orifice onwards were irradiated with step intervals between 25 μm and 5 mm. The position of orifice was determined by exposure of the device at various positions until no reflection of the inner capillary was measured. Ten frames with an exposure time of 0.10 s per frame were collected in SAXS and 30 frames were collected for USAXS.

Recorded two-dimensional (2D) scattering patterns are normalized to an absolute intensity scale (see SI S6). One-dimensional (1D) scattering patterns $I(q)$ were obtained by averaging 2D patterns over 360° for all SAXS measurements and for static measurements-only in USAXS according to standard procedures.[3] 2D time-resolved USAXS measurements were averaged over azimuthal angles between -20° and +20°. 1D patterns were averaged using all frames measured per measurement and corrected for background scattering using SAXSutilities2 software as described in SI sections S4 and S5.[4] Plotting and further data analysis has been done using OriginPro software.

Unless stated otherwise, all 1D scattering profiles and 2D scattering patterns shown throughout this paper are background corrected.

**Supporting Information 2            Fiber extrusion versus Fragment extrusion**

For analytical purposes and applications, bijels are continuously extruded as fibers (**Figure S2)**. Extrusion in microfluidic devices occurred inhomogeneously as fibers undulated, stuck to the glass and formed large accumulates in the capillaries (see SI video). These factors affect the time-resolved interpretation of bijel formation using the positional measurements. Alternatively, extruding fragments showed homogeneous movement throughout the capillary. As the length of the fiber was significantly larger than the diameter, we assumed that the phase separation kinetics are similar as in a fiber.

Fiber extrusions were typically performed using a precursor flowrate of 0.80 mL/h and 5.00 mL/h for toluene. Fragments are extruded using flowrates of 0.40 and 17.50 mL/h for, respectively, precursor and toluene. These fragments were extruded as a rate of ~30 s$^{-1}$ with an average spacing of 3 – 5 mm between the fragments.

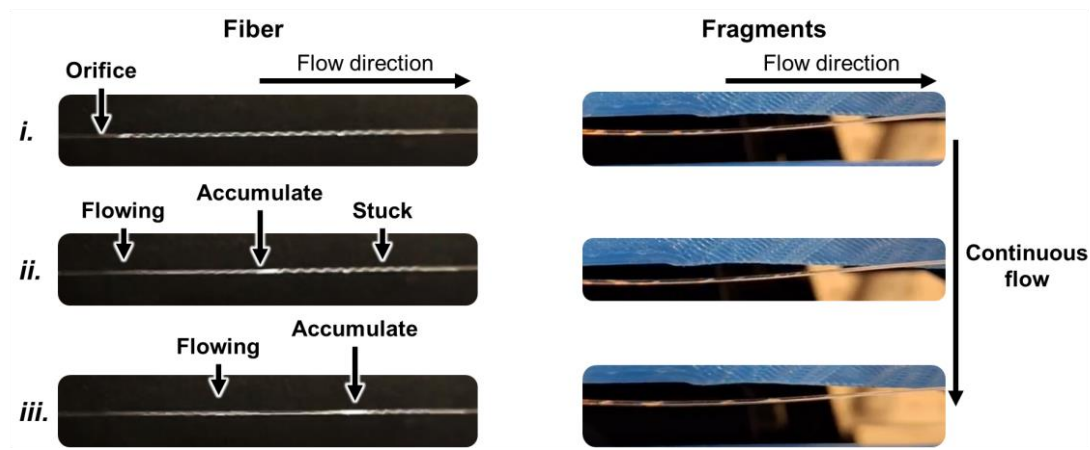

**Figure S2**: Time recordings of the extrusion of (left) fibers, and (right) fragments of bijel.

**Supporting Information 3**  **Velocity profile through capillary**

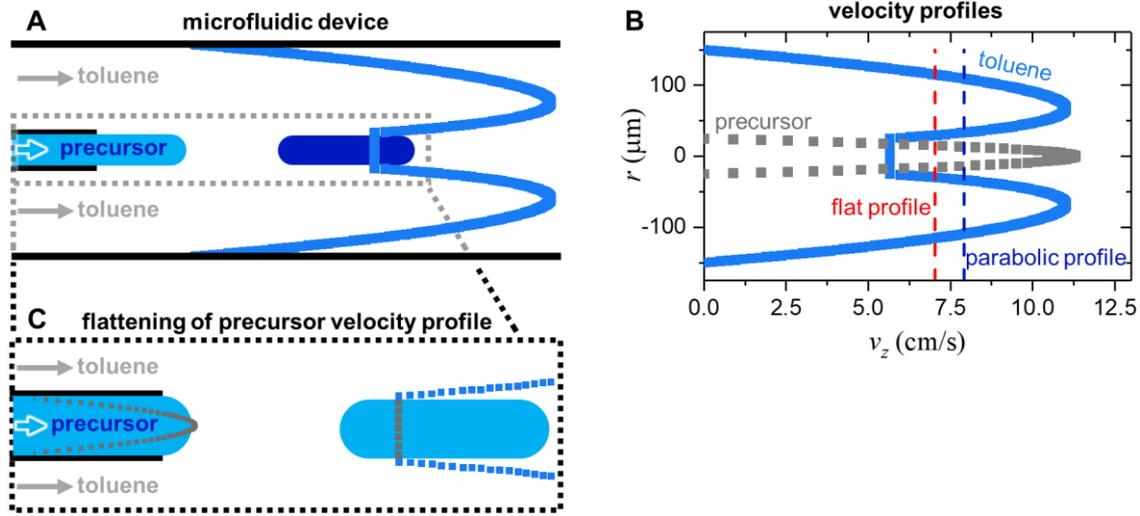

**Figure S3**: Velocity profiles in microfluidic device during fragment extrusion. **A)** Illustration of velocity profile in microfluidic device. **B)** Plot of velocity profiles against radial position *r* for (grey squares) precursor in 50 μm diameter inner capillary with $Q_{precursor}$ = 0.40 mL/hr, (blue squares) toluene in 300 μm diameter outer capillary with $Q_{toluene}$ = 17.50 mL/hr, (blue dashed line) averaged parabolic velocity and (red dashed line) flat profile. **C)** Proposed flattening of precursor velocity profile before and after injection.

The velocity of an extruded bijel fragment in a microfluidic device can be estimated assuming either a Laminar flow or constant profile in the outer capillary. The average velocities between both methods hardly differ as discussed next.

The Laminar velocity profile $v_z(r)$ can be calculated by treating the fragments as solid-like materials, as illustrated in **Figure S3A**. Using cylindrical coordinates and assuming no slip boundary conditions on the capillary walls ($v_z(r = R_2) = 0$) and fragment surface ($v_z(r = R_1 = U)$, $v_z(r)$ is calculated as:[5]

$$v_z(r) = -\frac{dP}{dz}\left(\frac{R_2^2 - r^2}{4\mu}\right) + \left(U + \frac{dP}{dz}\left(\frac{R_2^2 - R_1^2}{4\mu}\right)\right)\frac{\ln\left(\frac{R_2}{r}\right)}{\ln\left(\frac{R_2}{R_1}\right)}$$

with

$$\frac{dP}{dz} = \frac{2\mu \ln\left(\frac{R_2}{R_1}\right)}{\pi J}\left(Q_{tol} + \frac{2\pi U}{\ln\left(\frac{R_1}{R_2}\right)}\left[\frac{R_1^2}{2}\ln\left(\frac{R_1^2}{2}ln\left(\frac{R_1}{R_2}\right)\right) + \frac{R_2^2}{4} - \frac{R_1^2}{4}\right]\right)$$

and

$$J = \left(\frac{R_1^4}{4} - \frac{R_2^4}{4}\right)\ln\left(\frac{R_2}{R_1}\right) + \left(\frac{R_2^2}{2} - \frac{R_1^2}{2}\right)^2$$

and

$$U = \frac{Q_{\text{prec}}}{\pi R_1^2}$$

where $R_1$ is the fiber's radius, $R_2$ the outer capillary's radius, $\mu$ the viscosity of toluene, $U$ the translational velocity of the fragment, $Q_{\text{prec}}$ and $Q_{\text{tol}}$ the flow rate of, respectively, precursor and toluene.

As the fragment slightly undulates in the capillary, the average fiber velocity has been calculated via

$$v_{z,average} = \frac{1}{2R_2} \int_{-R_2}^{R_2} v_z(r) \mathrm{d}r$$

Using the experimental conditions of $R_1 = 50$ μm, $R_2 = 150$ μm, $Q_{\text{toluene}} = 17.50$ mL/hr, $Q_{\text{precursor}} = 0.40$ mL/hr and $\mu_{\text{toluene}} = 0.59$ mPa·s (20 °C) gives $v_{z,average} = 7.9$ cm/s as indicated in **Figure S3B**.

The undulation of the fragment disrupts the velocity profile, affecting the actual value for $v_{z,average}$. Alternatively, a flat velocity profile has been calculated via:

$$v_{fragment} = \frac{Q_{\text{precursor}} + Q_{\text{toluene}}}{\pi R_2^2}$$

Using the same experimental conditions gives $v_{fiber} = 7.0$ cm/s as indicated in **Figure S3B**. This shows that both methods gives similar average velocities. Using $v_{fragment}$ allows to compare phase separation kinetics across different experimental conditions while the actual time scale may slightly vary. In the main text, the time scales during extrusion have been estimated from the flat velocity profile.

The velocity profile of the precursor inside the 50 μm inner capillary can be calculated assuming $v_z(r = R) = 0$ i.e. no slip boundary condition at capillary wall via:[5]

$$v_z(r) = \frac{2}{\pi R^2} Q \left(1 - \left(\frac{r}{R}\right)^2\right)$$

The faster flow of toluene exerts a shear force on the fragment's outer surface upon extrusion, increasing the fragment's velocity. As a result, the velocity profile near the outer surface increases from 0 to $U$ (as described before) during extrusion. This may flatten the velocity profile in the fragment (**Figure S3C**).

**Supporting Information 4**     SAXS background subtraction and liquid scattering

Data subtraction on static and flowing samples have different background references. In static capillary samples, $I(q)$ of a precursor mixture contains scattering of silica TMA particles, CTAB, liquid mixture and quartz glass walls (**Figure S4A**). The effect of CTAB has been ignored as preliminary measurements showed no significant scattering due to the surfactant (not shown). The scattering of a quartz capillary filled with the liquid components of the precursor is significantly weaker than the precursor including silica nanoparticles. Scattering of only silica TMA particles has been calculated by subtracting the scattering of the quartz capillary filled with the liquid components of the precursor. The subtracted scattering of silica particles overlaps with the signal of the precursor. This shows that the precursor's scattering is dominated by silica nanoparticles independent of liquids and quartz glass capillary.

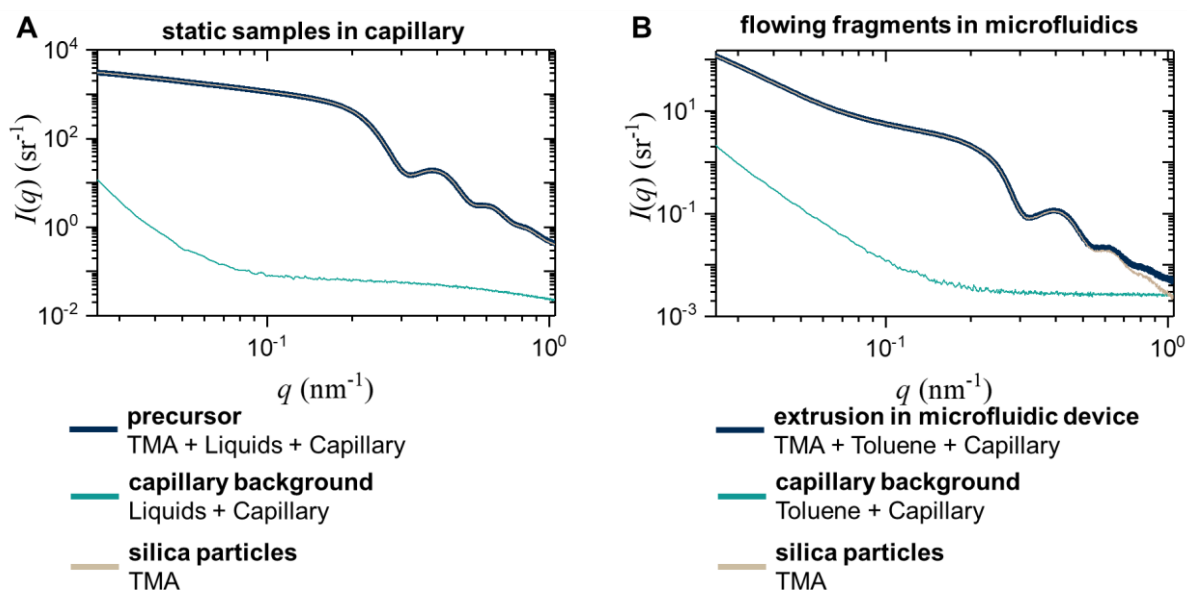

**Figure S4**: Background subtraction procedure for $I(q)$ for **A)** static samples in capillaries, and **B)** flowing samples in microfluidic devices.

Upon extrusion, the precursor's liquid composition changes due to the removal of 1-propanol. We assume the scattering of the bijel fragment originates solely from the nanoparticles due to weak scattering of the liquids. Therefore, the measured $I(q)$ during extrusion consists of the scattering of nanoparticles, toluene and the borosilicate glass capillary (**Figure S4B**). Subtraction of the scattering of a microfluidic capillary solely filled by toluene results in the scattering of the silica nanoparticles. The subtracted $I(q)$ mostly overlap with the unsubtracted scattering; limited deviations for $q > 0.7$ nm$^{-1}$ are observed. This range has been neglected as it solely corresponds to the intraparticle structure i.e. no interparticle structural contributions.

**Supporting Information 5**  USAXS background subtraction and data reduction

The measured scattering profiles during extrusions have a large contribution of the background for $q < 10^{-2}$ nm$^{-1}$. A microfluidic device filled with toluene has been subtracted from the measured scattering using SAXSutilities2 / 2D tools (**Figure S5A**). For better representation of the 2D patterns, the horizontal shadow region between the individual detectors has been patched by mirroring (caving) the images along the horizontal and vertical planes (**Figure S5Aiv**).

Data reduction of the corrected data is done by integrating between azimuthal angles of -20° and 20° to characterize anisotropic features in the 2D patterns. **Figure S5B** shows reduced 1D $I(q, 54\text{ ms})$ profiles integrated over 360° or azimuthal angles between -20° and 20°. An obvious difference in $I(q, 54\text{ ms})$ can be seen for $q < 10^{-2}$ nm$^{-1}$: a shoulder in $I(q, 54\text{ms})$ is present in the measured data, whereas a maximum is observed at $q = 10^{-2}$ nm$^{-1}$ in the corrected data.

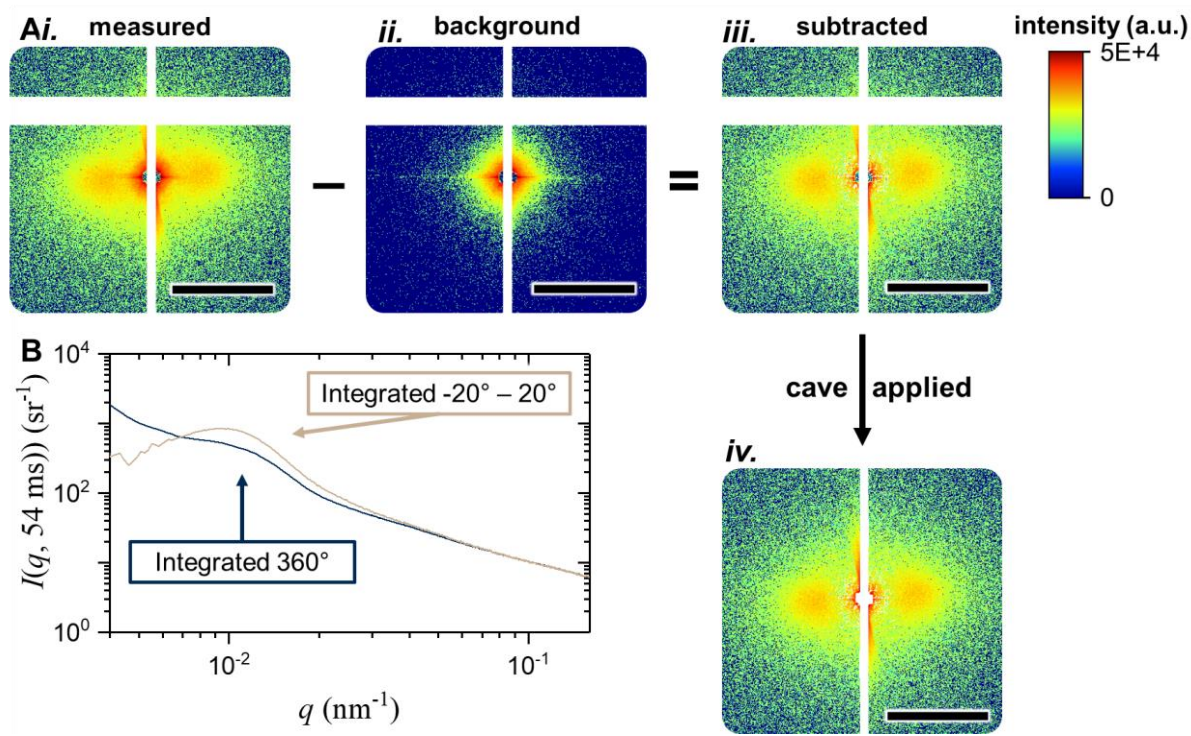

**Figure S5**: Background correction procedure of measured USAXS data from an extruded bijel fragment after 54 ms of extrusion. **A)** Different steps in background correction of 2D scattering patterns. Scale bar is 0.010 nm$^{-1}$. **B)** Comparison between 1D $I(q, 54$ ms$)$ for integration of (blue) 360°, and (sand) azimuthal -20°-+20°.

**Supporting Information 6**           **Normalization of measured $I(q, t_i)$**

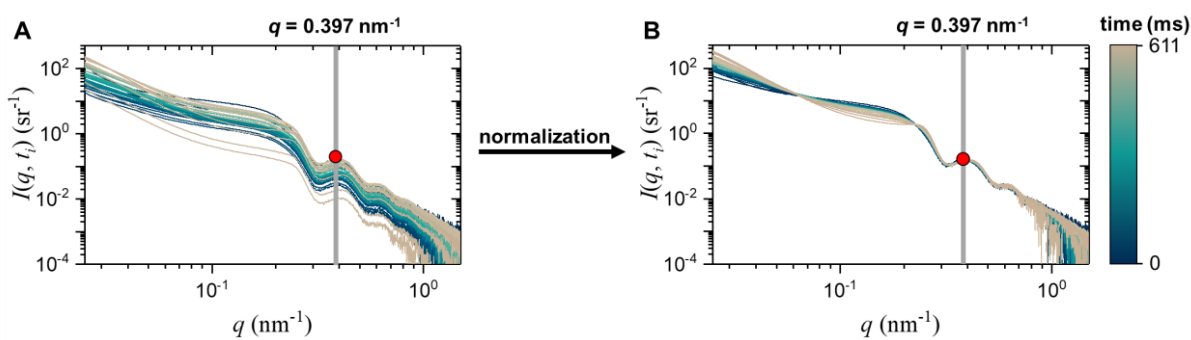

**Figure S6**: <u>Normalization procedure of scattered intensity profiles in SAXS.</u> **A)** $I(q, t_i)$ acquired during extrusion. **B)** Normalized $I(q, t_i)$ by using $I_i(q = 0.397 \text{ nm}^{-1}) = I_0(0.397 \text{ nm}^{-1})$.

The number of irradiated nanoparticles varies per measurement due to practical effects, resulting in an off-set change in $I(q, t_i)$ as shown in **Figure S6A**. To account for this variation, normalization has been done for the scattering of an individual nanoparticles by setting $I(q = 0.397 \text{ nm}^{-1}, t_i)$ equal to $I(q = 0.397 \text{ nm}^{-1}, 0 \text{ ms})$ as shown in **Figure S1B**. The scattering for $q > 0.30$ nm$^{-1}$ solely originates from silica nanoparticles and the scattering profiles after renormalization closely fall on top of each other. Thus, this renormalization allows to compare the different $I(q, t_i)$ profiles also at lower $q$-values.

**Supporting Information 7**          SAXS Form factor of aqueous Ludox TMA nanoparticles

**Figure S7** shows the scattering profiles of aqueous Ludox TMA dispersions acidified to pH 1.8 at various weight fractions measured in sealed quartz capillaries (diameter 1.5 mm, wall thickness 10 µm). A background scattering of water at pH 1.8 has been subtracted from the measured scattering profiles. The form factor for Ludox TMA, $P_{TMA}(q)$, is set as $I(q)$ for $w_i$ = 0.005 $g_{SiO2}$/$g_{dispersion}$. This is the lowest concentration with a proper signal-to-noise ratio and without structural development for $q < 0.3$ nm$^{-1}$.

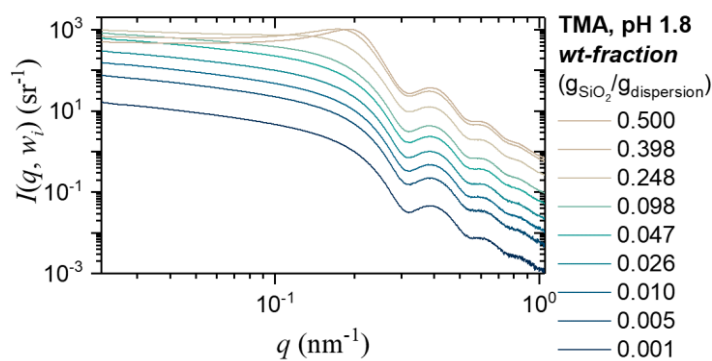

**Figure S7**: $I(q, w_i)$ of aqueous Ludox TMA dispersions at various wt-fractions.

**Supporting Information 8**          Calculating SAXS structure factor during bijel extrusion

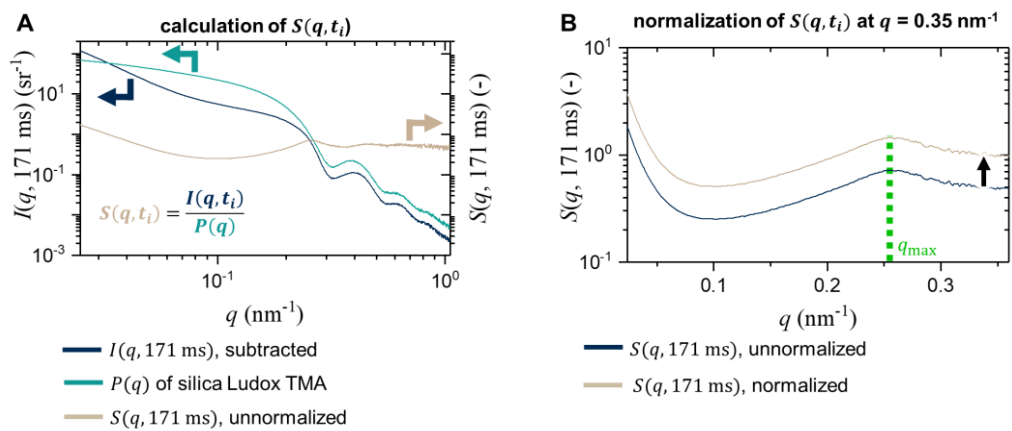

**Figure S8**: Calculating $S(q, t_i)$ under SAXS conditions. **A)** Conversion of $I(q, 171\text{ ms})$ into $S(q, 171\text{ms})$ for bijel extrusion after 171 ms. **B)** Normalization of $S(q, 171\text{ ms})$ using $S(q = 0.35\text{ nm}^{-1}, t_i) = 1$ with indication of $q_{max}$ (green dashed line).

The structure factor is extracted using the factorization of $S(q, t_i) = \frac{I(q,t_i)}{P(q)}$. **Figure S8A** shows the division of $I(q, t_i)$ of a bijel fragment after 171 ms of extrusion by the form factor $P(q)$ of silica TMA nanoparticles (SI S7) corresponding to scattering originating from individual particles without interactions.. As the number of silica particles within the beam is impossible to quantify, $S(q, t_i)$ profiles are normalized using $S(q = 0.35\text{ nm}^{-1}, t_i) = 1$ (**Figure S8B**) as signal originates from particles.

$S(q, t_i)$ often contains a maximum correlation peak located at $q_{max}$ at $q > 0.2$. Assuming that Bragg relation is valid, the mean particle distance (measured from core to core) is the reciprocal of the peak position given by (**Figure S8B**) by $d = \frac{2\pi}{q_{max}}$, corresponding to 24 nm i.e. particles are closely packed.

**Supporting Information 9**          Simulation of $I(q, t_i)$ and $S(q, t_i)$ within two-phase model

Bijel formation is proposed to occur via a two-phase model system describing the irreversible attachment of nanoparticles to the liquid-liquid interface. As phase-separation occurs radially inward into the cylindrical fragment, the measured scattered intensity $I_{meas}(q, t)$, can be described as the superposition of the scattering intensity of free and attached nanoparticles via (**Figure 2E** in main text):

$$I_{meas}(q, t) = \phi(t)I_1(q) + (1 - \phi(t))I_2(q)$$

where $I_1(q)$ is the scattering intensity of freely dispersed particles, $I_2(q)$ of attached nanoparticles and $\phi(t)$ the time-dependent volume fraction of free nanoparticles.

We describe $\phi(t)$ via $\phi(t) = A \exp(-at)$. Using boundary conditions $\phi(0) = \phi_0$ and $\phi(t_f) = \phi_f$, where $\phi_0$ and $\phi_f$ are, respectively, the initial and final volume fraction of freely dispersed nanoparticles gives as expression $\phi(t) = \phi_0 \left( \exp \frac{\ln\left(\frac{\phi_0}{\phi_f}\right)}{t_f} t \right)$. $t_f$ has been determined from the experimental measurement when no significant change in $I(q)$ occurs for $q < 0.3$ nm$^{-1}$.

$I_1(q)$ and $I_2(q)$ are measured experimentally for, resp., the static precursor mixture and extruded bijel fragment. The structure factors $S(q, t_i)$ are calculated using the form factor of silica TMA (see SI S7).

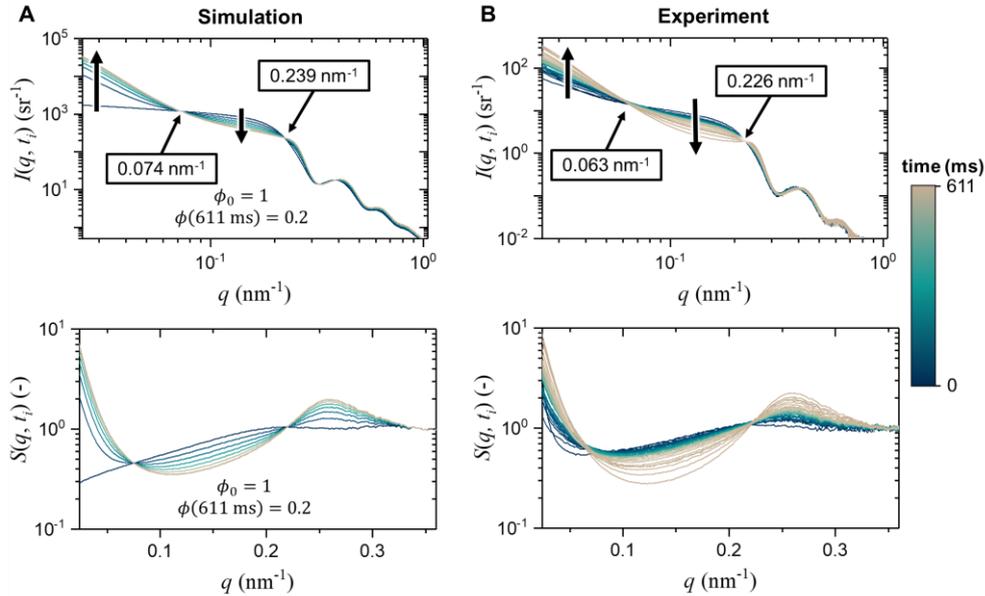

**Figure S9A:** $I(q, t_i)$ and $S(q, t_i)$ patterns as **A)** simulated, and **B)** experimentally measured.

The model has been tested by calculating $I(q, t_i)$ at various times. In these simulations, $\phi_0$ equals 1 i.e. all particles in the precursor are freely dispersed. $t_f$ was set to 611 ms i.e. the latest time recorded. Here, we first set $\phi_f = 0.2$ as guess-function.

**Figure S9A** plots the simulated and measured $I(q, t_i)$ and $S(q, t_i)$ using the model described above. Comparing both methods shows similar patterns in both $I(q, t_i)$ and $S(q, t_i)$. Both results have two iso-scattering points near $q = 0.07$ nm$^{-1}$ and $q = 0.23$ nm$^{-1}$. The patterns show that $I(q, t_i)$ and $S(q, t_i)$ remain constant for $q > 0.3$ nm$^{-1}$, decreases between the iso-scattering points and increases for $q < 0.074$ nm$^{-1}$.

Despite these similar patterns, the simulation has three main differences compared to the measurement: I) larger $q$-values for the iso-scattering points, II) less decrease in $I(q, t_i)$ and $S(q, t_i)$ between the iso-scattering points, and III) $S(q, 0\text{ ms})$ decreases for $q < 0.07$ nm$^{-1}$ which increases in the experiment. These differences may arise due to the value selected for $\phi_f$. To study the effect of $\phi_f$, the simulations were adjusted by varying $\phi_f$ between 0.01 and 0.5.

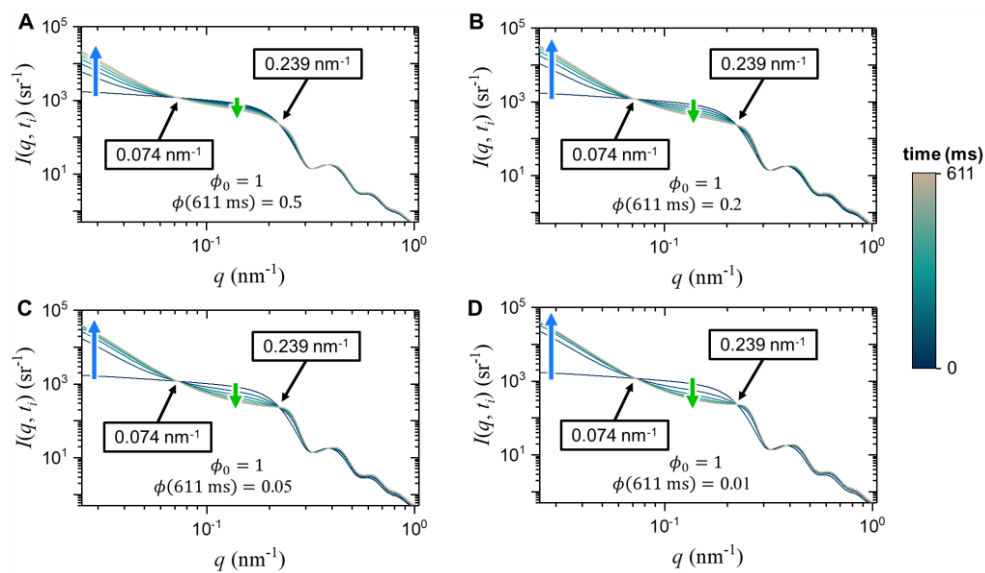

**Figure S9B**: Effect of $\phi_f$ on $I(q, t_i)$ using $\phi_0 = 1$ for $\phi_f$ equals **A)** 0.5, **B)** 0.2, **C)** 0.05, and **D)** 0.01.

**Figure S9B** shows four simulated $I(q, t_i)$ patterns with various $\phi_f$ while keeping $\phi_0 = 1$. Interestingly, the iso-scattering points remain at $q = 0.074$ and $q = 0.239$ nm$^{-1}$ for various $\phi_f$. However, the change in $I(q, t_i)$ highly depends on $\phi_f$. Lower values of $\phi_f$ result in a larger decrease in $I(q, t_i)$ for $q$ ranging between the iso-scattering points as shown by the green arrows. Simultaneously, $I(q, t_i)$ increases more rapidly for $q < 0.074$ nm$^{-1}$ indicated by the blue arrows. These results show that $\phi_f$ only affects the kinetics of the bijel formation without altering the position of the iso-scattering points and $I_{meas}(q, 0)$.

The differences in iso-scattering points and $S(q, 0)$ between the simulation and experiment, though, may originate from the differences in scattering of the precursor. In the simulation, $I_1(q)$ has been set using a static precursor which is actually flowing in the experiment. Despite these differences, no further optimization of the model has been done at this stage. This proposed model easily describes the similar pattern in both $I(q, t_i)$ and $S(q, t_i)$ compared to the experiment. Therefore, our model can describe the phase separation during bijel formation. However, the scale this model applies, has not been reported yet.

To validate the scale of the kinetic model, the normalized structure factor at $q = 0.025$ nm$^{-1}$ (corresponding to clusters of 250 nm) and 0.150 nm$^{-1}$ (individual nanoparticles) are calculated for various values of $\phi_f$ and $\phi_0 = 1$. **Figure S9C** plots $S(q = 0.025 \text{ nm}^{-1})$ and $S(q = 0.150 \text{ nm}^{-1})$ as measured experimentally and simulated for various $\phi_f$. It shows for $q = 0.025$ nm$^{-1}$ that no distinct value for $\phi_f$ matches the experimental structure factors. For $q = 0.150$ nm$^{-1}$, the simulation and experiment resemble each other for $\phi_f = 0.10$. The structure factors at both scales indicate that the two-phase model simulates the kinetics on the nanoparticle-level i.e. attachment of nanoparticles on the liquid-liquid interface. Larger structures like aggregates and pores, cannot be described by this model since other factors like vanderWaals-forces and coarsening of the domains deforms the particle scaffold.

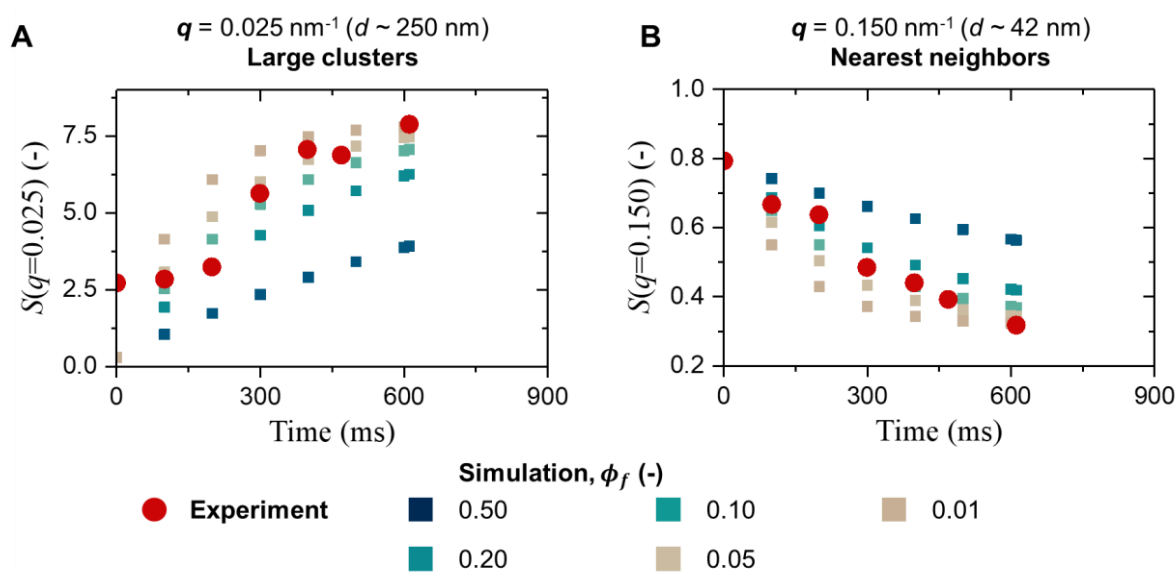

**Figure S9C**: Comparison between structure factors determined experimentally (circle symbols) and simulated with various values for $\phi_f$ (square symbols) for **A)** $q = 0.025$ nm$^{-1}$ and **B)** $q = 0.150$ nm$^{-1}$.

**Supporting Information 10**  Scattering intensity of flowing and static precursor

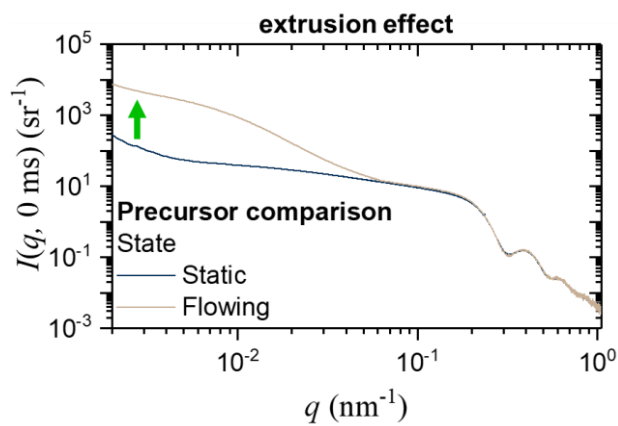

**Figure S10:** $I(q, 0 \text{ ms})$ of the precursor mixture as measured in (black) static capillary i.e. no flow, and (sand) flowing immediately after injection in a coaxial flow of toluene in a microfluidic device.

**Figure S10** shows $I(q, 0\text{ ms})$ of the precursor mixture under static and flowing conditions. The scattering intensity for the precursor under static and flowing conditions is similar for $q > 0.06$ nm$^{-1}$, while its scattering is significantly stronger for $q < 0.06$ nm$^{-1}$ under flowing conditions. This deviation may be explained by an extrusion effect which mechanism remains unclear in this study.

**Supporting Information 11**      **Calculation of Scattering Length Densities (SLD) and their effect on the contrast in SAXS and USAXS.**

To illustrate the importance of different contributions to SAXS and USAXS intensity, simple theoretical estimates were made. First of all, the typical scattering contrast was estimated as shown in **Table S11**. First of all, the real part of the refractive index $n = 1 - \delta$ of substances constituting the solution was obtained from reference [6], where $\delta$ is the refractive index decrement. Then, the scattering length density (SLD) values were calculated using

$$\rho_i = 2\pi\delta/\lambda^2$$

for several cases. Note that the exact compositions of the two liquid media, which are undergoing the spinodal phase decomposition, are time-dependent, approaching the oily phase mostly dominated by toluene and the aqueous phase, which is believed to be mostly composed of water and glycerol. In addition, silica nanoparticles with a much higher SLD, are dispersed in the system.

The scattering intensity is scaled with the square of the SLD contrast, $(\rho_i - \rho_j)^2$ between media $i$ and $j$. For the liquid phases, it reaches its maximum of $(\rho_{aq} - \rho_t)^2$ at the end of the phase separation. While the liquid phases have relatively low contrast, silica nanoparticles have much higher SLD. As a result, the scattering due to silica nanoparticles exceeds the scattering intensity due to the density difference between the fluids, by a factor of about 30:

$$(\rho_{SiO_2} - \rho_t)^2 \approx (\rho_{SiO_2} - \rho_{aq})^2 \approx (\rho_{SiO_2} - \rho_p)^2 \approx 30\,(\rho_{aq} - \rho_t)^2$$

As a result, the signal in the SAXS measurements is dominated by silica nanoparticles. However, this is no longer the case in the USAXS data, which correspond to much larger spatial scales. For order-of-magnitude estimations, the following model was used.

For the USAXS regime, one can neglect details on the scale of the particle size and approximate the particle layer by a uniform layer as illustrated in the **Figure S11A**. Let us consider a liquid droplet of the radius $R$ and the SLD $\rho_2$. It is covered by a silica shell with a thickness $d$ and SLD $\rho_3$. This core-shell particle is then inserted in another liquid with SLD $\rho_1$. The scattering intensity $I(q)$ (= differential scattering cross-section) of such a core-shell particle can be calculated as

$$I(q) = (4\pi)^2 \left[ (\rho_3 - \rho_1) \frac{\sin q(R+d) - q(R+d)\cos q(R+d)}{q^3} - (\rho_3 - \rho_2) \frac{\sin qR - qR\cos qR}{q^3} \right]^2$$

**Table S11**: Material properties and calculated SLD of precursor components and toluene.

|  | vol-% (v/v) | density (g mL⁻¹) | $\delta$ at 12.4 keV (-) | SLD (10⁻⁴ nm⁻²) |
|---|---|---|---|---|
| **Precursor** | | | | |
| DEP | 7.8 | 1.12 | $1.59 \times 10^{-6}$ | 10.0 |
| H$_2$O | 43.5 | 1.00 | $1.50 \times 10^{-6}$ | 9.44 |
| 1-propanol | 38.1 | 0.803 | $1.23 \times 10^{-6}$ | 7.72 |
| glycerol | 10.6 | 1.261 | $1.85 \times 10^{-6}$ | 11.6 |
| Total | 100 | 0.96 | | $\rho_p = 9.06$ |
| **Toluene** | 100 | 0.8623 | $1.28 \times 10^{-6}$ | $\rho_t = 8.02$ |
| **Water+glycerol** | 80 + 20 | 1.05 | | $\rho_{aq} = 9.87$ |
| **Silica** | 100 | 2.2 | $2.99 \times 10^{-6}$ | $\rho_{SiO_2} = 18.8$ |

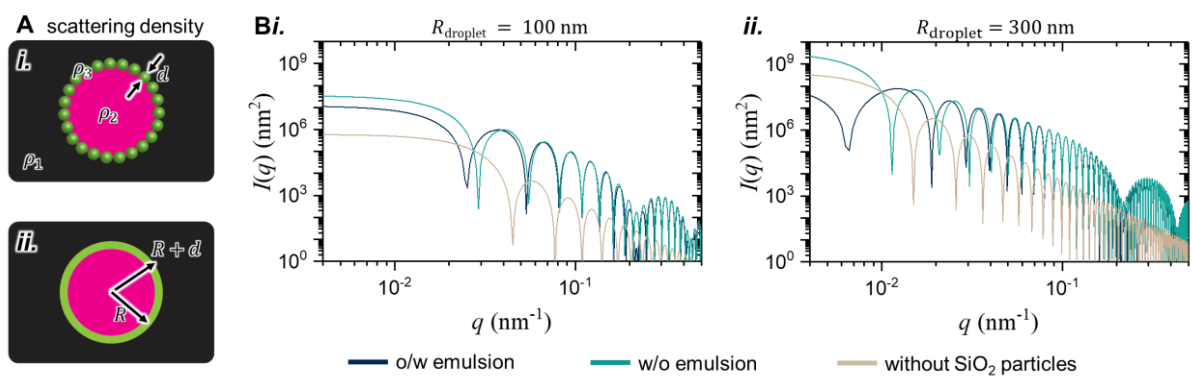

**Figure S11**: Scattering cross sections. **A*i*)** Schematic droplet of liquid phases 1 and 2 separated by a solid particle phase 3 with thickness $d$ all with respective scattering length density $\rho_i$. **A*ii*)** Schematic dimensions of core-shell droplets with liquid radius $R$ and droplet radius $R + d$. **B)** Scattering intensity profiles of core-shell droplets against scattering vector for droplets with radius of *i*) 100 nm and *ii*) 300 nm.

The scattering profiles $I(q)$ are presented for no silica shell (sand), water-in-oil (green), and oil-in-water droplets (blue) as shown in **Figure S11B**. For $R = 100$ nm, the scattering of a liquid droplet without the silica shell is much weaker (by a factor of 20 to 60 at the smallest $q$-values) than that for droplets with silica although some difference can be already seen between water-in-oil and oil-in-water droplets due to interference between the liquid-liquid and silica-liquid contrasts. For $R = 300$ nm (i.e., droplet diameter of 600 nm is about 20 times larger than the nanoparticle diameter of 29 nm), the scattering cross-section due to the liquid-liquid contrast is comparable to the scattering due to the silica shell. A much stronger interference effect between the two types of contrast can also be seen.

Let us consider a 1D model consisting of periodically-arranged sets of 4 layers: watery phase – silica – oily phase – silica. The thickness of the stack is $L = L_w + d + L_o + d = L_w + L_o + 2d$. The SLD are given above. The scattering amplitude is proportional to the Fourier amplitudes of the structure

$$A_n \propto \int_{-L/2}^{L/2} e^{inq_0 x} \rho(x)\, dx,$$

where $q_0 = 2\pi/L$ is the main spatial frequency of the structure and $n$ is the order of the Fourier component. By placing the origin of the $x$-axis in the middle of the oily layer and subtracting $\rho_w$ from all the contrasts, one gets

$$A_n \propto \int_{-L_o/2}^{L_o/2} (\rho_o - \rho_w) \cos nq_0 x\, dx + 2\int_{L_o/2}^{L_o/2+d} (\rho_{SiO_2} - \rho_w) \cos nq_0 x\, dx$$

$$= \frac{2}{nq_0}\left[(\rho_o - \rho_w)\sin n\pi \frac{L_w}{L} + (\rho_{SiO_2} - \rho_w)\left(\sin n\pi \frac{L_w+d}{L} - \sin n\pi \frac{L_w}{L}\right)\right].$$

Finally, we assume a para-crystal model with the scattering intensity

$$I(q) \propto \sum_n |A_n|^2 \frac{1}{\pi} \frac{nw}{(q-nq_0)^2 + (nw)^2},$$

where $w$ is the half-width at half-maximum of the first Lorentzian diffraction peak. Note that the width of the higher-order reflections is assumed to grow with $n$. This assumption is different from the one usually used to describe Bragg peak broadening in a crystal powder due to the finite-size effect. The increasing peak width assumption used here corresponds to the "second-type disorder" in the terminology of Guinier[7], which corresponds to a gradual loss of the positional order. The predictions of this para-crystal model is illustrated in **Figure S11C** below.

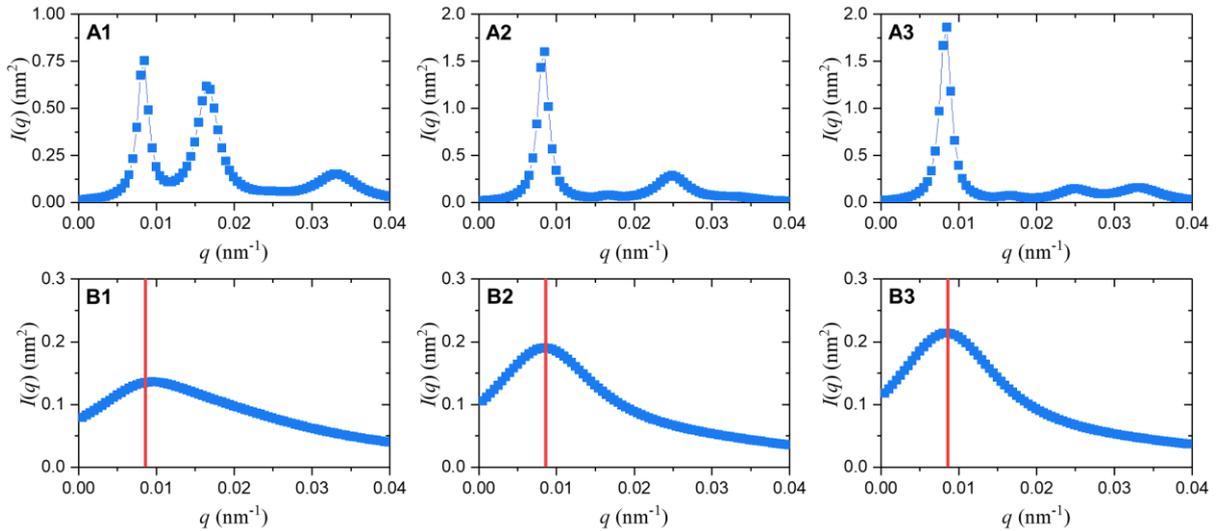

**Figure S11C**: Prediction of the para-crystal model for $w = 0.1\, q_0$ (top row of An panels) and $w = q_0$ (bottom row of Bn panels). $L_w = L_o$ (panels A1 and B1), $L_w = 2L_o$ (panels A2 and B2), and $L_w = 3L_o$ (panels A3 and B3). In all cases $L_w + L_o = 700$ nm. The vertical red line corresponds to $q_0 = 2\pi/L$.

To better illustrate the importance of different terms in $I(q)$, the results are presented for $w = 0.1\, q_0$ so that the peaks can be seen separately. In panel A1, $L_w$ is assumed to be the same as $L_o$ so that the distance between silica layers is exactly one-half of the whole structure period $L = L_w + L_o + 2L_{SiO_2}$. In this case, the silica layers can only contribute to even diffraction peaks and the first peak at $q_0 = 2\pi/L$ originates solely from the water-oil scattering contrast. However, it is not anymore true for $L_w \neq L_o$. Moreover, one can see that the relative intensity of the peaks depends on the interference between different contributions which, in turn, depend on the exact geometry assumed in the model.

We now turn to a much more realistic situation with $w = q_0$ since the bicontinuous structure is not really periodic but can only be characterized by a characteristic sizes of the pores of two types. Here the peaks greatly overlap forming a single broad peak.

So, we can conclude that the scattering of the nanoparticles dominates the SAXS scattering intensity recorded. This is confirmed by the visibility of oscillations in the $I(q)$ profiles in the main text. However, in USAXS the scattering induced by the contrast between the oil- and aqueous phase start to contribute to similar extents as silica nanoparticles.